\documentclass[12pt]{article}

\usepackage{graphicx}
\usepackage{amsmath,amssymb}
\usepackage{bm}
\usepackage{graphicx, color}
\usepackage{wrapfig}
\begin{document}
\title{
\begin{flushright}
\ \\*[-80pt] 
\begin{minipage}{0.2\linewidth}
\normalsize
IPMU16-0007 \\*[50pt]
\end{minipage}
\end{flushright}
{\Large \bf 
Occam's Razor in Quark Mass Matrices
\\*[20pt]}}

\author{ 
\centerline{
~Morimitsu~Tanimoto$^{1}$ \quad and \quad Tsutomu T. ~Yanagida$^{2}$}
\\*[20pt]
\centerline{
\begin{minipage}{\linewidth}
\begin{center}
$^1${\it \normalsize
Department of Physics, Niigata University,~Niigata 950-2181, Japan} \\
$^2${\it \normalsize
Kavli IPMU, TODIAS, University of Tokyo, Kashiwa 277-8583, Japan} 
\end{center}
\end{minipage}}
\\*[50pt]}
\date{
\centerline{\small \bf Abstract}
\begin{minipage}{0.9\linewidth}
\medskip 
\medskip 
\small
On the standpoint of  Occam's Razor approach,  we consider the minimum number of parameters in the quark mass matrices needed  for the successful CKM mixing and CP violation.
We impose three zeros in the down-quark mass matrix with taking the diagonal  up-quark mass matrix to reduce the number of free parameters.  The three zeros are maximal zeros in order to have a CP violating phase in the quark mass matrix. Then,  there remain six real parameters and one CP violating phase,
which is the minimal number to reproduce the observed data of the down-quark masses and the CKM parameters.
The twenty textures with three zeros are examined.
 Among them,   the  thirteen textures  are viable for the down-quark mass matrix.
 As  a representative of these textures, we  discuss a texture $M_{d}^{(1)}$
in details.  By using the experimental data on $\sin 2\beta$, $\theta_{13}$ and $\theta_{23}$ together with
the observed quark masses, 
the Cabibbo angle is predicted to be close to the experimental data.
 It is found that this surprising result remains unchanged in all other viable textures.
 We  also investigate  the correlations among $|V_{ub}/V_{cb}|$, $\sin 2\beta$
 and $J_{CP}$.
 For all textures, the maximal value of the ratio $|V_{ub}/V_{cb}|$ is $0.09$, which is smaller than
 the upper-bound of  the experimental data, $0.094$.
 We hope that this prediction will be tested in future experiments.
\end{minipage}
}

\begin{titlepage}
\maketitle
\thispagestyle{empty}
\end{titlepage}

\section{Introduction}

The standard model is now well established by the recent discovery of the Higgs boson.
In spite of the success of the standard model, underlying physics determining the 
quark and lepton mass matrices is still unknown. Because of this there have been
proposed a number of models based on flavor  symmetries, but no convincing
model has been proposed.

A long times ago, Weinberg \cite{Weinberg:1977hb} considered a mass matrix for the down-quark sector in the basis of the up-quark mass matrix being diagonal.  He assumed a vanishing (1,1) element in the $2\times2$ matrix and imposed a symmetric form of the matrix in order to reduce the number of free parameters. Then, the number of free parameters is reduced to only two and hence he succeeded to predict the Cabibbo angle  to be  $\sqrt{m_d/m_s}$, which is very successful and  called as
the   Gatto, Sartori, Tonin relation \cite{Gatto:1968ss}. \footnote{
Fritzsch extended the above approach to the three
family case \cite{Fritzsch:1977vd,Fritzsch:1979zq}. 
He  set four zeros in each down-quark  and up-quark mass matrices, which are both
symmetric.  Then, there were eight parameters against the  ten observed data.
However, it was ruled out by the observed CKM element $V_{cb}$.
Ramond, Roberts and Ross  also presented the systematic work with  four or five zeros
  for the symmetric or hermitian quark mass matrix \cite{Ramond:1993kv}.
Their textures are also not viable under the precise experimental data at present,
because four or five zeros is too tight to reproduce the  ten observed data.}

The success of the Weinberg approach encouraged many authors to consider various flavor symmetries in the quark matrices
by extending it to the three family case. In this paper, however, we point out that the Cabibbo angle is predicted successfully
in the framework of ``Occam's Razor'' approach proposed in the lepton sector \cite{Harigaya:2012bw}.
\footnote{ The Occam's Razor approach predicts the CP-violating phase $\delta =\pm  \pi/2$ in the neutrino oscillation. It is very much interesting that one of the predictions, 
$\delta \simeq -\pi/2$, is favored in a global analysis of the neutrino oscillation data
 \cite{CP}.}  In other word, we show that
the Weinberg matrix can be obtained without any symmetry.

On the standpoint of Occam's Razor approach,  we consider the minimum number of parameters needed  for the successful CKM mixing and CP violation without assuming the  symmetric or hermitian  mass matrix of down quarks. 
We impose three zeros in the down-quark mass matrix. We always take the up-quark mass matrix to be diagonal since we do not need any off-diagonal element to explain the observation. 
Therefore,  the down-quark mass matrix  is given with  six complex parameters.
Among them,  five phases can be removed by the phase redefinition of the three right-handed and three left-handed down quark fields. After the field-phase rotation, there remain
 six real parameters and one CP violating phase,
which are the minimal number to reproduce the seven  observed data, that is the three down-quark masses and the four CKM parameters.
It is emphasized that the three-zero texture keeps one CP violating phase in the down-quark mass matrix.
In the present Occam's Razor approach with  three families, we show that the successful prediction of the Cabibbo angle is obtained. It is surprising that the Weinberg's anzatz, that  is $(1,2)$ and $(2,1)$  elements to be  symmetric, is derived
in our Occam's Razor approach.

In order to reproduce the bottom-quark mass and the CKM mixing angles, $V_{us}$ and  $V_{cb}$, we take the elements $(3,3)$, $(2,3)$,  $(1,2)$ of the $3\times 3$ down-quark mass matrix to be non-vanishing.
Then, we have $_6 C_3=20$ textures with three zeros. 
For our  convenience, we classify them in two categories;
 (A) where a $(2,2)$ element is non-vanishing
 and (B)  where  a $(2,2)$ element is vanishing.
 In the category (A),   the  six textures with three zeros are viable for the down-quark mass matrix.
 In the category (B), there are seven viable textures with three zeros. 
 Finally, we have found that  those thirteen textures are all consistent with  the  present experimental data
 on quark masses and the CKM parameters. However, we should note that
 some textures are equivalent each other due to the freedom of the unitary transformation  of the  right-handed down quarks.

  In section 2, we show viable   down-quark mass matrices with three zeros in the categories  (A) and (B),
   and explain how to obtain the CKM mixing angles and the CP violating phase.
   In section 3, we show numerical results for those mass matrices.
   The discussion and summary is devoted in section 4.
   In Appendices A and B, we show  unfavored down-quark mass matrices
    and the above redundancy of our textures, respectively.


\section{Quark mass matrix in the  Occam's Razor approach}

\subsection{Three texture zeros for down-quarks}

Let us discuss the down-quark mass matrix. We always take the basis where the up-quark mass matrix is diagonal.
The number of free parameters in the down-quark mass matrix is reduced by putting zero at several elements in the matrix.
We consider the three texture zeros, which provide the minimum number of parameters needed  for the successful CKM mixing and CP violation. We never assume  any flavor symmetry. We call this as ``Occam's Razor `` approach.
 
Before investigating the quark mass matrix, we present our setup in more details.
The Lagrangian for the quark Yukawa sector is given by

\begin{align}
\mathcal{L}_Y&=y_{\alpha\beta}^u \bar Q_{L\alpha} u_{R\beta} \tilde h+
y_{\alpha\beta}^d \bar Q_{L\alpha} d_{R\beta} h \ ,
\label{lagrangian}
\end{align}
where  $Q_{L\alpha}$,  $u_{R\beta}$,  $d_{R\beta}$ and $h$ denote
the left-handed quark doublets, the right-handed up-quark singlet,
the right-handed down-quark singlet, and  the Higgs doublet, respectively.
The quark mass matrices are given as  $m_{\alpha\beta}=y_{\alpha\beta} v_H$ with $v_H=174.104$ GeV.
In order to reproduce the observed  quark masses and the CKM matrix
with the minimal number of parameters, 
 we take the diagonal basis in the up-quark sector,
\begin{equation}
M_u=
\begin{pmatrix}
m_u & 0 & 0 \\
0 & m_c& 0 \\
0 & 0 & m_t
\end{pmatrix}_{LR} \ .
\label{up}
\end{equation}
For the down-quark mass matrix, we impose the three texture zeros.
Then, the texture of the down-quark mass matrix $M_d$ is given with  six complex parameters.
The five phases can be removed by the phase rotation of the three right-handed and three left-handed
down-quark fields. Therefore, there remains six real parameters and one CP violating phase,
which are the minimal number to reproduce the observed data of masses and the CKM parameters.

Now, we can discuss  textures for the down-quark mass matrix.
Let us  start with taking $(3,3)$, $(2,3)$, $(1,2)$ elements of $M_d$ to be non-vanishing values
 to reproduce the observed bottom quark mass and the CKM mixing angles, $V_{us}$ and
$V_{cb}$.
\footnote{By  the unitary transformation of the right-handed quarks, we can move
to textures with other non-vanishing entries.}
Then, we have $_6C_3=20$ textures with three zeros for the down-quark mass matrix.
For our convenience, we classify them in two categories, (A) and (B), as explained in the introduction.
 In (A),  we have  $_5 C_2=10$ textures with a non-vanishing $(2,2)$ element
 and in  (B)  we have also ten textures with a vanishing  $(2,2)$ element.
 
We first discuss  textures in the category (A).
The ten textures are written as follows:
  
\begin{align}
&
\begin{pmatrix}
0 & A & 0 \\
A' & B& C \\
0 & C' & D
\end{pmatrix}  , \quad
\begin{pmatrix}
A' & A & 0 \\
0 & B& C \\
0 & C' & D
\end{pmatrix}  , \quad
\begin{pmatrix}
0 & A & 0 \\
0 & B& C \\
A' & C' & D
\end{pmatrix}  , \quad
\begin{pmatrix}
0 & A & C' \\
A' & B& C \\
0 & 0 & D
\end{pmatrix}  , \quad
\begin{pmatrix}
A' & A & C' \\
0 & B& C \\
0 & 0 & D
\end{pmatrix}  ,
 \nonumber\\ 
\nonumber\\
&
\begin{pmatrix}
0 & A & C' \\
0 & B& C \\
A' & 0 & D
\end{pmatrix}  , \quad
\begin{pmatrix}
0 & A & 0 \\
A' & B& C \\
C' & 0 & D
\end{pmatrix}  , \quad
\begin{pmatrix}
A' & A & 0 \\
C' & B& C \\
0 & 0 & D
\end{pmatrix} , \quad
\begin{pmatrix}
0 & A & C' \\
0 & B& C \\
0 & A' & D
\end{pmatrix}  , \quad
\begin{pmatrix}
A' & A & 0 \\
0 & B& C \\
C' & 0 & D
\end{pmatrix}  ,
\label{massmatrixA}
\end{align}
where $A, A', B, C, C'$ and $D$ are complex parameters. 
We will show that the first six textures are consistent with  the  present experimental data.
Those  six down-quark mass matrices  are parametrized 
after removing five phases  by the phase rotation of quark fields as follows:
\begin{align}
&M_d^{(1)}=
\begin{pmatrix}
0 & a & 0 \\
a' & b \ e^{-i\phi}& c \\
0 & c' & d
\end{pmatrix}_{LR} \ , \quad
M_{d}^{(2)}=
\begin{pmatrix}
a' & a & 0 \\
0 & b \ e^{-i\phi}& c \\
0 & c' & d
\end{pmatrix}_{LR} \ , \quad
M_{d}^{(3)}=
\begin{pmatrix}
0 & a & 0 \\
0 & b \ e^{-i\phi}& c \\
a' & c' & d
\end{pmatrix}_{LR} \ , \nonumber\\
\nonumber\\
& M_{d}^{(4)}=
\begin{pmatrix}
0 & a & c' \\
a' &  b \ e^{-i\phi}& c \\
0 & 0 & d
\end{pmatrix}_{LR} \ , \quad
M_{d}^{(5)}=
\begin{pmatrix}
a' & a  & c' \\
0 & b \ e^{-i\phi}& c \\
0 & 0 & d
\end{pmatrix}_{LR} \ , \quad
M_{d}^{(6)}=
\begin{pmatrix}
0 & a & c' \\
0 & b \ e^{-i\phi}& c \\
a' & 0 & d
\end{pmatrix}_{LR} \ , 
\label{downmassmatrixA}
\end{align}
where $a, a', b, c, c'$ and $d$ are real parameters, and $\phi$ is the CP violating phase.
It should be stressed that our matrices are not symmetric at all. 
The CP violating phase  $\phi$ is put in the $(2,2)$ entry.
\footnote{One can put $\phi$ in other places
without changing the predicted mass eigenvalues and  CKM  elements
since the rephasing is possible in these mass matrices.}
In the next section, we examine those seven parameters numerically  to reproduce the three down quark masses, three CKM mixing angles and one CP violating phase.

We discuss briefly why the last four textures (seventh-tenth ones) in Eq.(\ref{massmatrixA}) are excluded by the experimental data.   The seventh and eighth textures in Eq.(\ref{massmatrixA}) give
 a vanishing CKM mixing angle.
 The ninth one gives us  one massless quark because the first column is a zero vector in the flavor space.
The last one cannot reproduce the magnitude of  the CP violation.
The details are shown in Appendix A.


 Now, we  discuss textures in the category (B), in which the $(2,2)$ element is zero.
 We can write the ten textures as follows:
  
\begin{align}
&
\begin{pmatrix}
A'& A & B \\
0 & 0& C \\
0 & C' & D
\end{pmatrix}  , \quad
\begin{pmatrix}
0 & A &B \\
A' & 0& C \\
0 & C' & D
\end{pmatrix}  , \quad
\begin{pmatrix}
0 & A & B \\
0 &0& C \\
A' & C' & D
\end{pmatrix}  , \quad
\begin{pmatrix}
A & A' & C' \\
B & 0& C \\
0 & 0 & D
\end{pmatrix}  , \quad
\begin{pmatrix}
A & A' & B \\
0 & 0& C \\
C' & 0 & D
\end{pmatrix}  , \nonumber\\
\nonumber\\
&
\begin{pmatrix}
0 & A & B \\
A' & 0& C \\
C' & 0& D
\end{pmatrix}  , \quad
\begin{pmatrix}
A & A' & 0 \\
B & 0& C \\
C' & 0 & D
\end{pmatrix}  ,  \quad
\begin{pmatrix}
A' & A & 0 \\
0 & 0& C \\
B & C' & D
\end{pmatrix}  , \quad
\begin{pmatrix}
0 & A & 0 \\
A' &0& C \\
B & C' & D
\end{pmatrix} , \quad
\begin{pmatrix}
A' & A & 0 \\
B &0& C \\
0 & C' & D
\end{pmatrix} .
\label{massmatrixB}
\end{align}
The first seven textures are also consistent with the present experimental data.
After removing five phases  by the phase rotation of 
quark fields, those  seven down-quark mass matrices  are parametrized as:
\begin{align}
&M_d^{(11)}=
\begin{pmatrix}
a' &a \ e^{-i\phi} & b \\
0 & 0 \ & c \\
0 & c' & d
\end{pmatrix}_{LR}  , \quad
M_{d}^{(12)}=
\begin{pmatrix}
0 & a \ e^{-i\phi} & b \\
a'& 0& c \\
0 & c' & d
\end{pmatrix}_{LR}  , \quad
M_{d}^{(13)}=
\begin{pmatrix}
0 & a \ e^{-i\phi} & b \\
0 & 0 & c \\
a' & c' & d
\end{pmatrix}_{LR}  , \nonumber\\
\nonumber\\
&M_{d}^{(14)}=
\begin{pmatrix}
a\ e^{i\phi} & a'  & c' \\
b & 0 & c \\
0 & 0 & d
\end{pmatrix}_{LR}  ,  \quad
M_{d}^{(15)}=
\begin{pmatrix}
 a \ e^{-i\phi}& a' & b \\
0 & 0 & c \\
c' & 0 & d
\end{pmatrix}_{LR} , \nonumber\\
\nonumber\\
&M_{d}^{(16)}=
\begin{pmatrix}
0  & a & b \\
a' & 0 & c \ e^{-i\phi}\\
c' & 0 & d
\end{pmatrix}_{LR} ,  \quad
M_{d}^{(17)}=
\begin{pmatrix}
a & a' & 0 \\
b & 0& c \ e^{i\phi}\\
c' & 0 & d
\end{pmatrix}_{LR}  . 
\label{downmassmatrixB}
\end{align}

The last three textures (eighth-tenth ones) in Eq.(\ref{massmatrixB}) are also excluded by the experimental data.   The  eighth and ninth textures in Eq.(\ref{massmatrixB}) give
 a vanishing  CKM  mixing angle.
The tenth ones  cannot reproduce the  magnitudes of the CP violation.
The details are discussed  in  Appendix A.

Finally, we comment on the freedoms of the unitary transformation of the  right-handed quarks. Since the CKM matrix is the  flavor mixing among  the left-handed quarks, 
 some  textures in Eqs.(\ref{downmassmatrixA}) and (\ref{downmassmatrixB})
are equivalent each other due to the freedom of the unitary transformation  of the  right-handed quarks. 
We show the redundancy among them in Appendix B.

\subsection{CKM parameters for  $M_d^{(1)}$}

Let us  show how to predict the CKM mixing angles  and the CP violation
taking  the case of $M_d^{(1)}$ in  Eq.(\ref{downmassmatrixA}) as a representative. 
Since the up-type quark mass matrix is diagonal, the CKM matrix is obtained
by diagonalizing the down-quark mass matrix $M_d^{(1)}$ .
 In order to determine the left-handed quark mixing angles, we study $M_d^{(1)} M_d^{(1)\dagger}$;
\begin{equation}
M_d^{(1)} M_d^{(1)\dagger}=
\begin{pmatrix}
a^2 & ab\  e^{i\phi} & ac' \\
ab\  e^{-i\phi} & a'^2+b^2+c^2 & bc'\  e^{-i\phi}+cd\\
ac' &  bc'\  e^{i\phi}+cd & c'^2+d^2
\end{pmatrix}_{LL} \ .
\label{MMdagger}
\end{equation}
By solving the eigenvalue equation of  $M_d^{(1)} M_d^{(1)\dagger}$,
we obtain
\begin{align}
&m_d^2+m_s^2+m_b^2=a^2+a'^2+b^2+c^2+c'^2+d^2 \ , \nonumber \\
&m_d^2 m_s^2+ m_s^ 2m_b^2+m_b^2 m_d^2= 
a^2 a'^2 +a^2 (c^2+ d^2)+a'^2 (c'^2+d^2) +c^2 c'^2+b^2 d^2 -2bcc'd\cos\phi\ ,   \nonumber \\
&m_d^2 m_s^2 m_b^2= a^2 a'^2 d^2 \ .
\label{massrelations}
\end{align}
On the other hand, the eigenvectors lead to the CKM  elements $V_{ij}$ and the CKM phase $\delta_{CP}$, which  is given in the PDG parametrization \cite{PDG}.
Those  are given in the leading order as follows:
\begin{equation}
|V_{us}|\simeq \frac{ab}{m_s^2}\left | \sin\frac{\phi}{2} \right | \ ,
\quad
|V_{cb}|\simeq \sqrt{2}\frac{c}{m_b} \left |\cos\frac{\phi}{2} \right |\ ,
\quad
|V_{ub}|\simeq \frac{a c'}{m_b^2} \ ,
\quad
\delta_{CP}\simeq \frac{1}{2}(\pi-\phi) \ ,
\label{CKM}
\end{equation}
where we adopt the approximate relations $b\sim c$ and $c'\sim d$,
which will be justified in our numerical results.

Although the source of the CP violation is only one, we have various measurements of the CP violating phase.
We can also define the other CP violating quantity, $\beta$ (or $\phi_1$),
which is the  one angle of the unitarity triangle:
\begin{equation}
\beta\ (\phi_1)=\arg \left ( -\frac{V_{cd}V_{cb}^*}{V_{td}V_{tb}^*} \right )  \ .
\end{equation}
Actually, $\sin 2\beta$ has been measured precisely in the B-factory \cite{PDG}.
It is given concisely in the leading order by
\begin{equation}
\sin 2\beta\simeq \sin \phi  \ .
\end{equation}
There is another CP violating observable, the Jarlskog invariant $J_{CP}$ \cite{Jarlskog:1985ht},
which is derived from the following relation:
\begin{align}
&iC\equiv [M_u M_u^\dagger, M_d M_d^\dagger]  \ , \nonumber \\
& \det C= -2 J_{CP} (m_t^2-m_c^2) (m_c^2-m_u^2) (m_u^2-m_t^2) (m_b^2-m_s^2) (m_s^2-m_d^2) (m_d^2-m_b^2) \ .
\label{Jcp}
\end{align}
The predicted one is exactly expressed in terms of the parameters of the mass matrix elements as:
\begin{align}
J_{CP}=\frac{-1}{(m_b^2-m_s^2) (m_s^2-m_d^2) (m_d^2-m_b^2) }\ j_{CP}  \ ,
\end{align}
where 
\begin{align}
j_{CP}={a^2 b c c' d \sin\phi} \ .
\end{align}

For $M_d^{(k)}\ (k=1-6)$ and $M_d^{(k)}\ (k=11-17)$, 
we summarize  three CKM  elements $V_{ij}$, the  CP violating phase $\delta_{CP}$  and
 $j_{CP}$ in Table 1.

\begin{table}[hbtp]
\begin{center}
\begin{tabular}{|l|c|c|c|c|c|}
\hline 
& $ |V_{us}|$ & $ |V_{cb}|$ &$ |V_{ub}|$& $\delta_{CP}$& $ j_{CP}$ \\
\hline 
&&&&&\\
  $M_{d}^{(1)}$, $M_{d}^{(2)}$,  $M_{d}^{(3)}$, $M_{d}^{(16)}$, $M_{d}^{(17)}$  
& $\frac{ab}{m_s^2} \left |\sin\frac{\phi}{2}\right |$  & 
$\frac{\sqrt{2}c}{m_b}\left |\cos\frac{\phi}{2}\right |$&  $\frac{a c'}{m_b^2}$& $\frac{1}{2}(\pi-\phi)$&  $a^2bcc' d \sin\phi$\\
&&&&&\\
\hline 
&&&&&\\
  $M_{d}^{(4)}$, $M_{d}^{(5)}$, $M_{d}^{(6)}$, $M_{d}^{(14)}$ & $\frac{ab}{m_s^2}$& $\frac{c}{m_b}$& $\frac{c'}{m_b}$& $\phi$&$a bcc' d^2 \sin\phi$\\
 &&&&&\\
\hline 
&&&&&\\
 $M_{d}^{(11)}$,$M_{d}^{(12)}$, $M_{d}^{(13)}$, $M_{d}^{(15)}$& $\frac{ac}{m_s^2}\frac{c'}{m_b}$& $\frac{c}{m_b}$& $\frac{b}{m_b}$& $\pi-\phi$&$a bc^2c' d \sin\phi$\\
 &&&&&\\
\hline 
\end{tabular}
\caption{The predicted CKM  elements, the CP phase and the CP violating measure $j_{CP}$,
where $|V_{ij}|$ and $\delta_{CP}$ are given in the leading order, and $j_{CP}$ is the exact one.}
\label{tab1}
\end{center}
\end{table}

\section{Numerical studies of Textures}
\subsection{Derivation of Weinberg's mass matrix for  $M_d^{(1)}$}
 Since the  texture of $M_d^{(1)}$ is the most familiar one among the thirteen textures 
in Eqs.(\ref{downmassmatrixA})
and  Eq.(\ref{downmassmatrixB}),
 we show how to obtain the numerical result by taking   $M_d^{(1)}$ as a typical representative.
 In order to reduce the number of the free parameters, we use the three
 observed down-quark masses as inputs. 
 We adopt the  data of the down-quark Yukawa couplings at the $M_Z$ scale
\cite{Antusch:2013jca} as
\begin{equation}
y_d=(1.58^{+0.23}_{-0.10}) \times 10^{-5}, \quad 
y_s=(3.12^{+0.17}_{-0.16}) \times 10^{-4}, \quad
y_b=(1.639\pm 0.015) \times 10^{-2},
\label{dyukawa}
\end{equation}
which give quark masses as $m_q=y_q v_H$ with $v_H=174.104$ GeV.
 For $M_d^{(1)}$, it is convenient to eliminate the parameters  $a'$, $d$ and $\phi$
by using Eq.(\ref{massrelations}), where the three quark masses
in Eq.(\ref{dyukawa}) are put at the $90\%$ C.L, respectively.
Then, there remain four parameters $a$, $b$,  $c$ and $c'$.
We calculate the CKM parameters by scanning these parameters  randomly
 in the region of $\{a,b,c,c',d\}=\{0, m_b\}$  with $a,b,c,c' \leq d$. 

  In fig.1, we show  the frequency distribution  of the predicted value of  $|V_{us}|$.
   The large angle is allowed as well as the small angle with almost same weight.
In fig.2, we show the $|V_{us}|$ versus $a/a'$. 
Since we do not assume any relation among the parameters  $a,b,c$ and  $c'$, 
the  larger values than $1/\sqrt{2}$ are allowed. 
 The observed Cabibbo angle favors around $a/a'=1$,
 which was assumed  in the Weinberg's approach.
 Since we do not assume $a/a'=1$, the allowed $|V_{us}|$ leis in $0-1$ at this stage. 
 We will show that the desired relation $a/a'\simeq 1$ is finally derived without using the observed Cabibbo angle  in our approach.
   
   In figs.3 and 4, we show the frequency distributions of   the predicted value of  $|V_{cb}|$ and $|V_{ub}|$, respectively. 
    The $|V_{cb}|$ is allowed up to the maximal mixing $1/\sqrt{2}$. On the other hand, the  $|V_{ub}|$ is predicted  to be smaller  than  $0.02$ because of the quark mass hierarchy
    as seen in Eqs.(\ref{massrelations}) and (\ref{CKM}).
However, it is still  much larger than the observed value.

\begin{figure}[h!]
\begin{minipage}[]{0.45\linewidth}
\includegraphics[width=8cm]{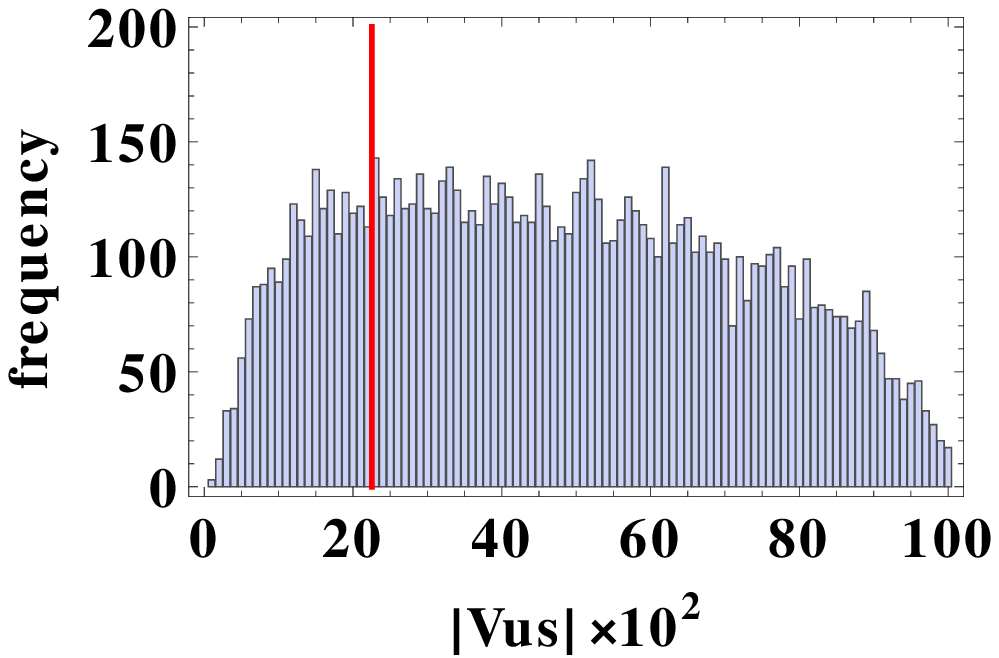}
\caption{The frequency distribution of the predicted  $|V_{us}|$   in arbitrary unit without the CKM constraints  for $M_{d}^{(1)}$. The red  line denotes the experimental value. }
\end{minipage}
\hspace{5mm}
\begin{minipage}[]{0.45\linewidth}
\includegraphics[width=8cm]{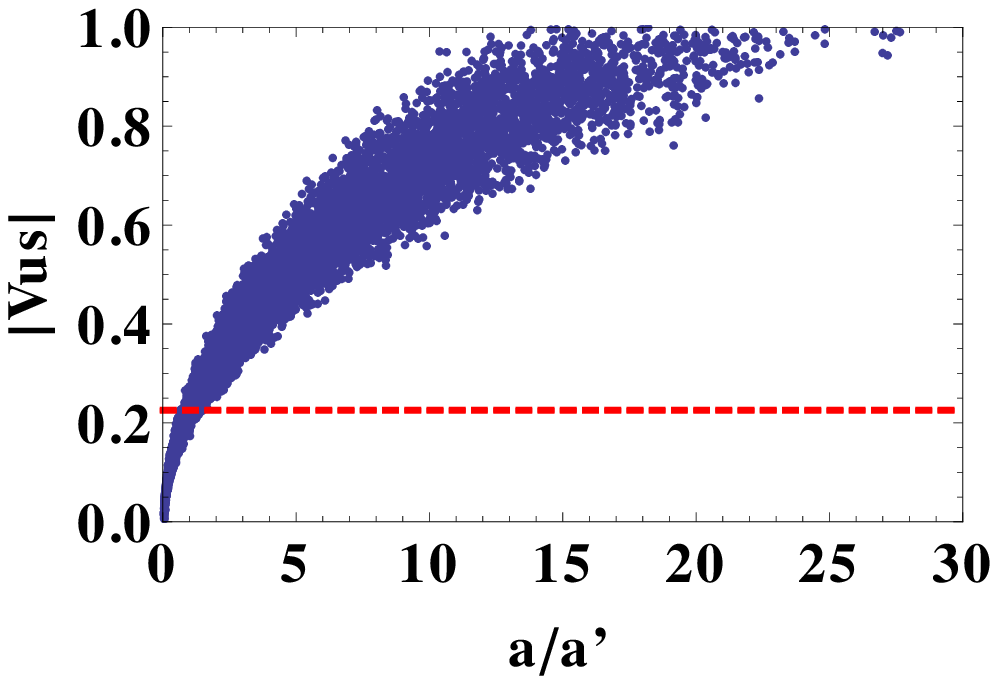}
\caption{The prediction of $|V_{us}|$ versus $a/a'$  without the CKM constraints for $M_{d}^{(1)}$.
The red dashed line denotes  the experimental value.}
\end{minipage}
\end{figure}
\begin{figure}[h!]
\begin{minipage}[]{0.45\linewidth}
\includegraphics[width=8cm]{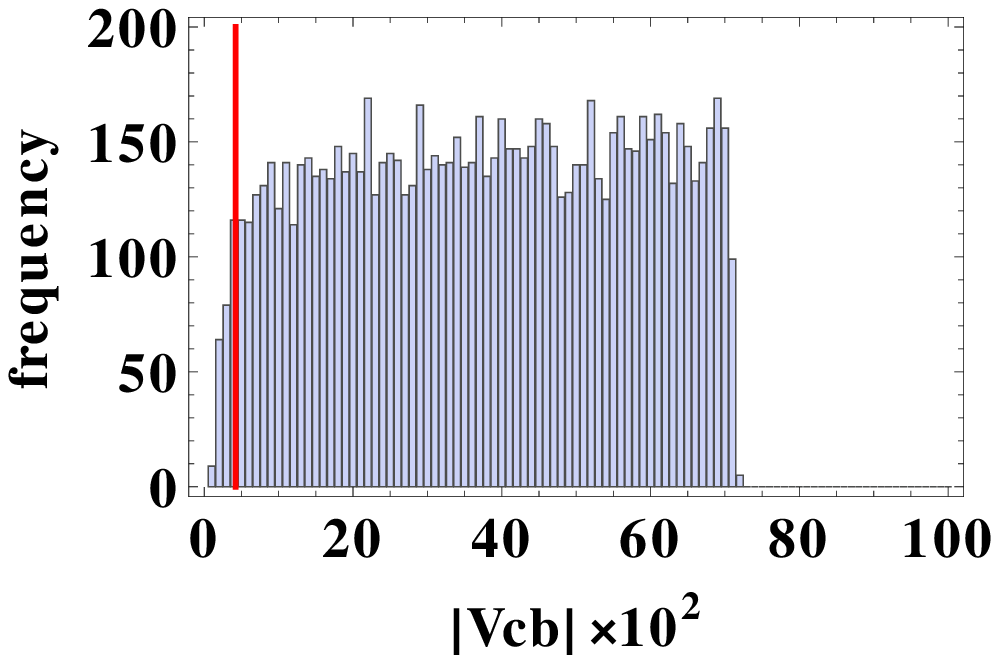}
\caption{The frequency distribution of the predicted  $|V_{cb}|$   in arbitrary unit without the CKM constraints  for $M_{d}^{(1)}$. The red  line denotes
 the  experimental value.}
\end{minipage}
\hspace{5mm}
\begin{minipage}[]{0.45\linewidth}
\includegraphics[width=8cm]{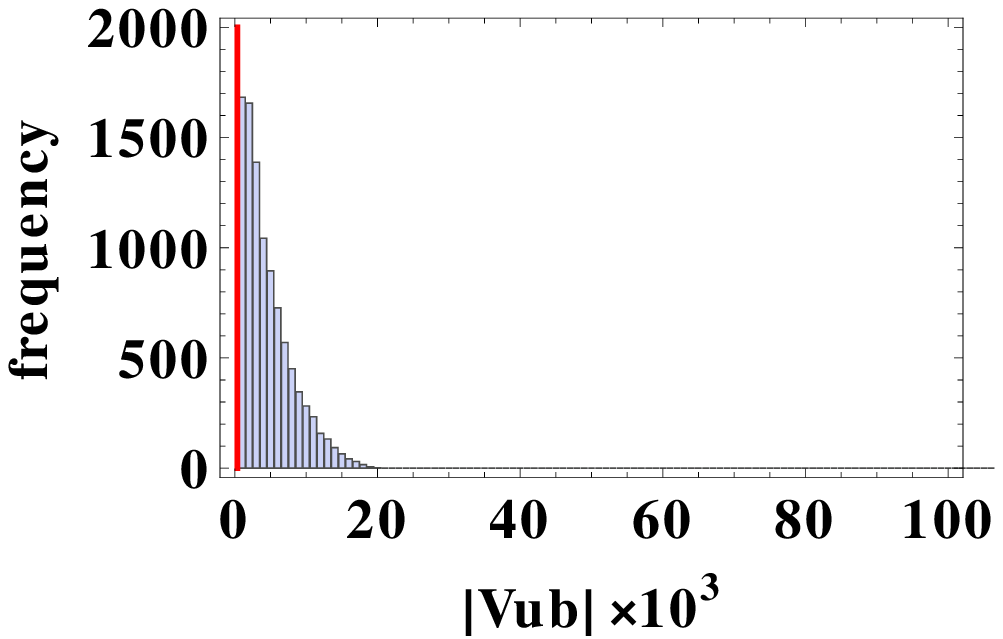}
\caption{The frequency distribution  of the predicted $|V_{ub}|$   in arbitrary unit without the CKM constraints 
for $M_{d}^{(1)}$. The  red  line denotes
 the  experimental value.}
\end{minipage}
\end{figure}

Here, we show the present experimental data on the  CKM mixing angles and the CP violating phase at the $M_Z$ scale \cite{Antusch:2013jca},
\begin{align}
&\theta_{12}=0.22735\pm 0.00072 \ , \quad\qquad\ \ \ 
\theta_{23}=(4.208\pm 0.064)\times 10^{-2},  \nonumber \\
&\theta_{13}=(3.64\pm 0.13)\times 10^{-3}, \quad\qquad
\delta_{CP}=1.208\pm 0.054\  [{\rm rad}] \ ,
\label{data}
\end{align}
where $\theta_{ij}$ is defined  in the  PDG parametrization \cite{PDG}.
We can also use the other CP violating measure, $\sin 2\beta$,  as an input parameter.
 The experimental data on  $\sin 2\beta$ is given by \cite{PDG}:
\begin{equation}
\sin 2\beta=0.682\pm 0.019 \ .
\label{data2}
\end{equation}
We will discuss the predicted $J_{CP}$ comparing with the experimental data,
\begin{equation}
J_{CP}=(3.06^{+0.21}_{-0.20})\times 10^{-5} \ , 
\end{equation}
which is obtained by the global fit of the CKM parameters \cite{PDG}.

\begin{figure}[h!]
\begin{minipage}[]{0.45\linewidth}
\includegraphics[width=8cm]{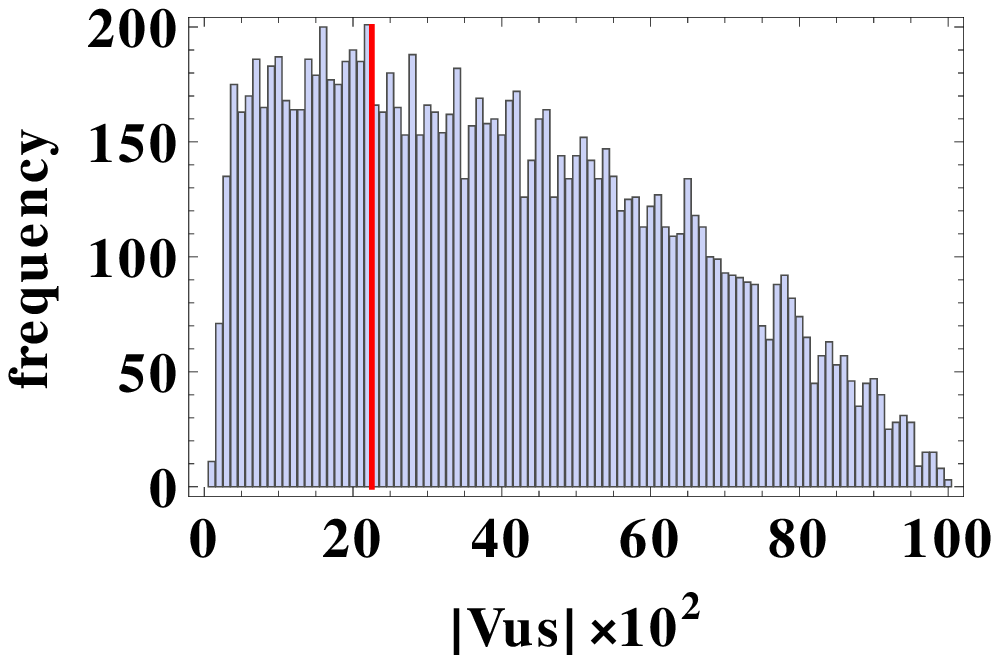}
\caption{The frequency distribution of $|V_{us}|$   in arbitrary unit with  the constraint of   $\sin 2\beta$  
for $M_{d}^{(1)}$. The red  line denotes the experimental value.}
\end{minipage}
\hspace{5mm}
\begin{minipage}[]{0.45\linewidth}
\includegraphics[width=8cm]{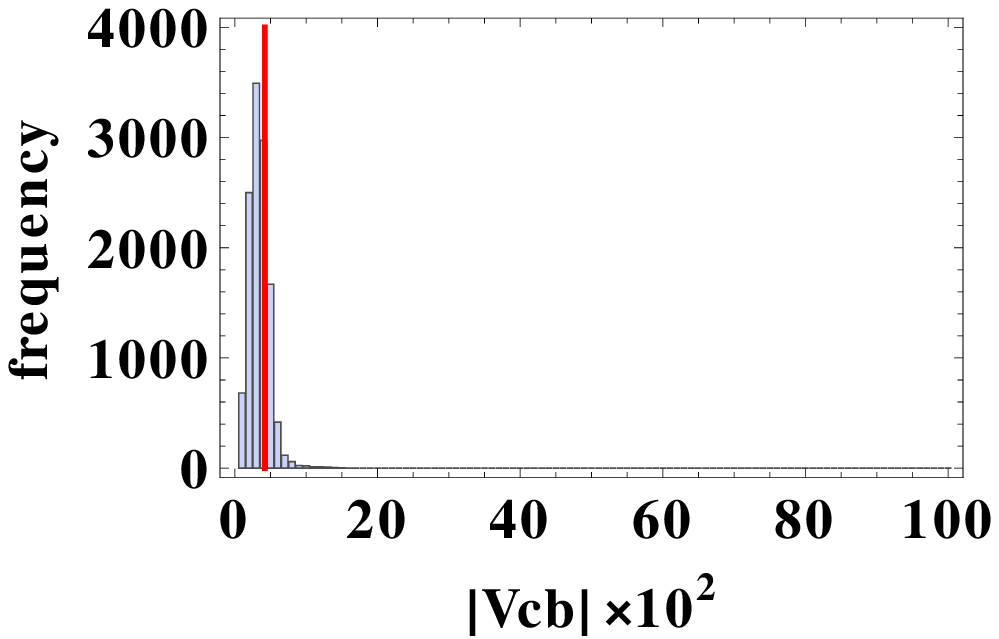}
\caption{The frequency distribution of $|V_{cb}|$   in arbitrary unit with  the constraint of   $\sin 2\beta$  
for $M_{d}^{(1)}$. The red  line denotes the experimental value.}
\end{minipage}
\end{figure}
\begin{wrapfigure}{r}{8cm}
\vspace{-5mm}
\includegraphics[width=7.5cm]{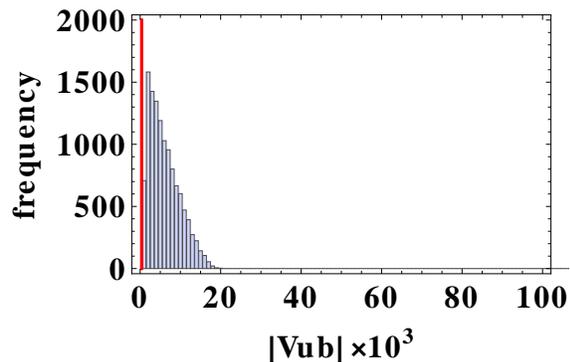}
\caption{The frequency distribution of $|V_{ub}|$   in arbitrary unit with  the constraint of  $\sin 2\beta$ for $M_{d}^{(1)}$. The red  line denotes the experimental value.}
\end{wrapfigure}

Let us, first, use the CP violating phase   in addition to
 the three quark masses. We adopt the observed data on $\sin 2\beta$  in Eq.(\ref{data2}),
 which is more sensitive to constrain the parameters compared with the data on $\delta_{CP}$.
 In figs.5 and 6,  we show the frequency distributions  of  the predicted values of  $|V_{us}|$ and $|V_{cb}|$, respectively. 
The predicted   $|V_{us}|$ is still broad  as $0-1$.
 However, the frequency distribution of $|V_{cb}|$ is impressive because the peak
is very sharp around the observed value. Thank to the constraint of
 the experimental data of $\sin 2\beta$,
 the predicted $|V_{cb}|$ is restricted to be in the narrow region at one push. 
In fig.7,  we show the  frequency distribution of  the predicted value of  $|V_{ub}|$.
This distribution is not much changed compared with the case in fig.4.


In the next step, we add the constraint  of the 
experimental data $\theta_{13}$ in Eq.(\ref{data}) .
In fig.8,   we show the  frequency distribution  for the predicted value of  $|V_{us}|$.
The peak of this distribution is around the experimental value although the large mixing is still allowed.
The frequency distribution  of  the predicted value of  $|V_{cb}|$
is presented in fig.9. 
This distribution is not much changed compared with the case in fig.6.

\begin{figure}[h!]
\begin{minipage}[]{0.45\linewidth}
\includegraphics[width=8cm]{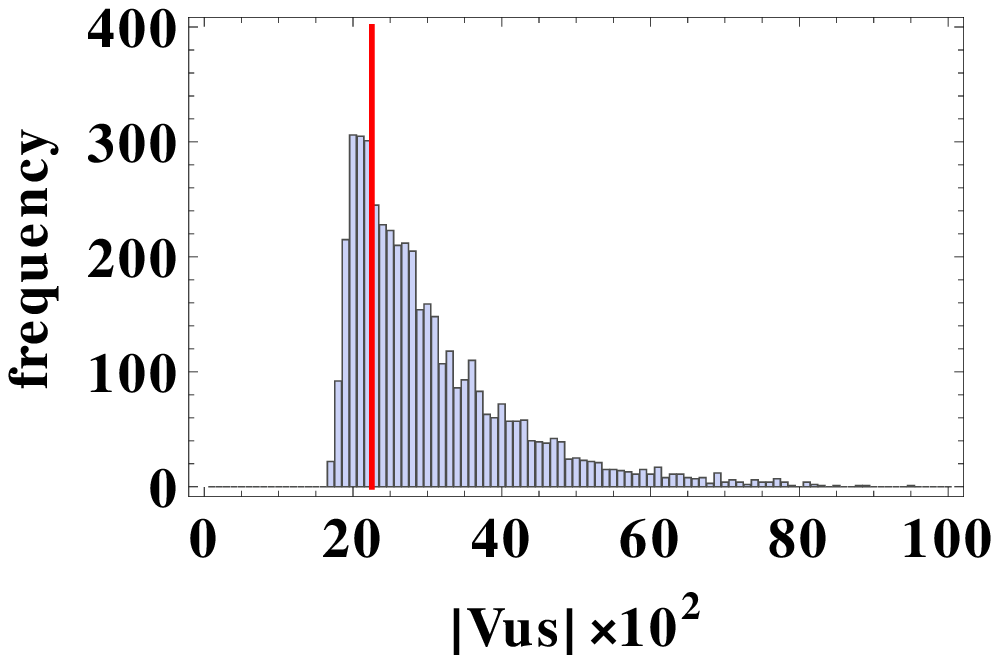}
\caption{The frequency of $|V_{us}|$   in arbitrary unit with  the constraints of
    $\sin 2\beta$ and $|V_{ub}|$ for $M_{d}^{(1)}$.
The red  line denotes the experimental value.}
\end{minipage}
\hspace{5mm}
\begin{minipage}[]{0.45\linewidth}
\includegraphics[width=8cm]{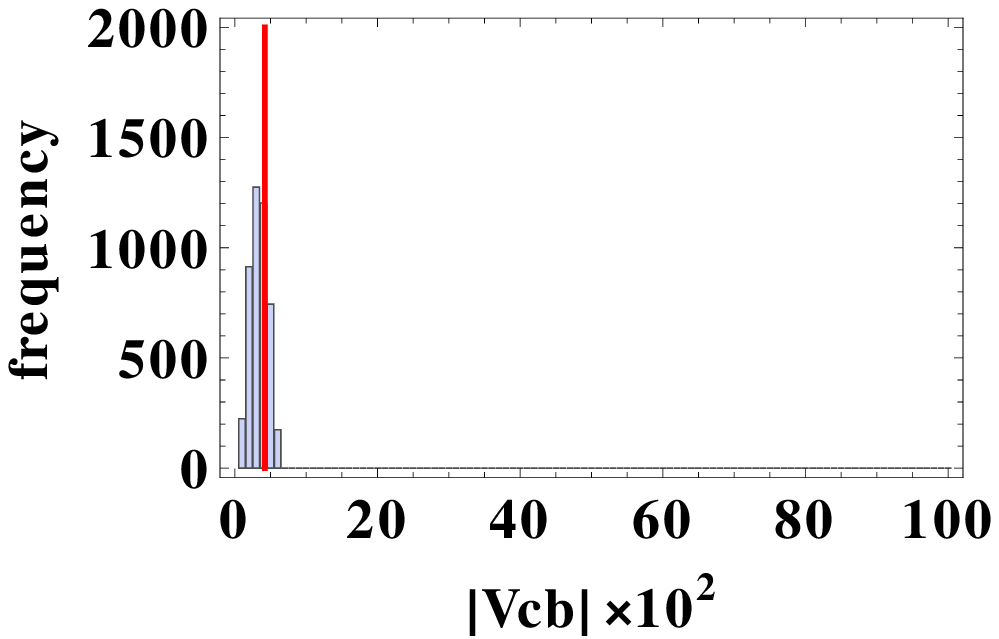}
\caption{The frequency of $|V_{cb}|$   in arbitrary unit with  the constraints of
    $\sin 2\beta$  and $|V_{ub}|$  for $M_{d}^{(1)}$.
The red  line denotes the experimental value.}
\end{minipage}
\end{figure}

At the last step,  we impose  the  constraint of the 
experimental data $\theta_{23}$ in Eq.(\ref{data}).
Then, we obtain the predicted  Cabibbo angle with a good accuracy.
In fig.10,  the  frequency distribution of  the predicted value of  $|V_{us}|$
is shown. The predicted  $|V_{us}|$ is shown  versus the parameter $a/a'$ in fig.11.
The prediction is completely consistent with the experimental data,
and the ratio  $a/a'=0.7-1.9$ is given.
Thus, the Weinberg's anzatz  $a/a'=1$ is obtained  by using
the experimental data $\sin 2\beta$, $\theta_{13}$ and $\theta_{23}$ without assuming 
any flavor symmetry.
Now, we can determine the parameter $a$ (or $a/a'$) precisely by using the experimental data
 $\theta_{12}$ in Eq.(\ref{data}).

\begin{figure}[h!]
\begin{minipage}[]{0.45\linewidth}
\includegraphics[width=8cm]{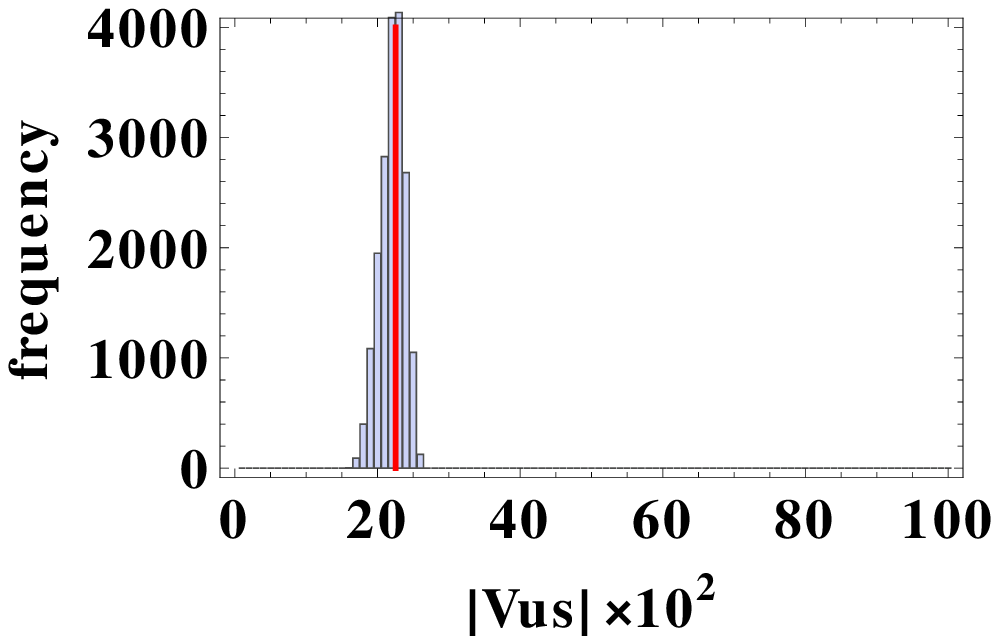}
\caption{The frequency of  $|V_{us}|$  in arbitrary unit with  the constraints of
  $\sin 2\beta$, $|V_{ub}|$ and   $|V_{cb}|$  for $M_{d}^{(1)}$.
The red  line denotes the experimental value.}
\end{minipage}
\hspace{5mm}
\begin{minipage}[]{0.45\linewidth}
\includegraphics[width=8cm]{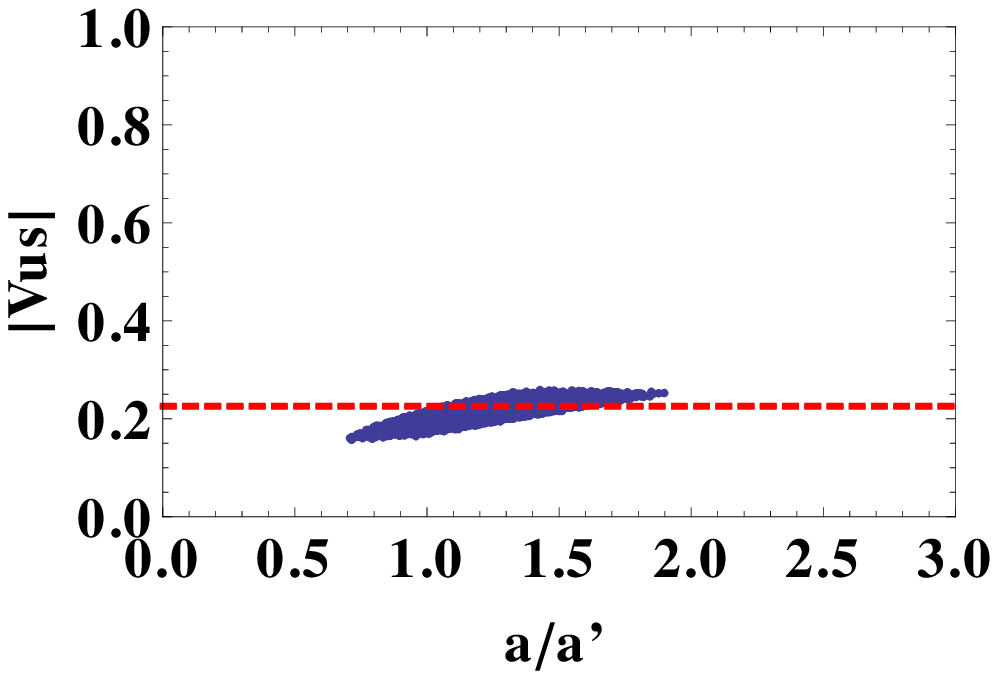}
\caption{The prediction of $|V_{us}|$ versus $a/a'$ with  the constraints of
  $\sin 2\beta$, $|V_{ub}|$ and  $|V_{cb}|$   for $M_{d}^{(1)}$.
The red dashed line denotes  the experimental value.}
\end{minipage}
\end{figure}

 The seven parameters of  $M_{d}^{(1)}$ are determined completely by the experimental data.
 However, we can add further investigations
 since  there is  still  large error-bars
 in the experimental data  $|V_{ub}|$.
  We show the  predicted ratio $|V_{ub}/V_{cb}|$ versus $\sin 2\beta$ in fig.12.
 The upper bound of the predicted ratio is $0.09$ while the experimental data allows up to
  $0.094$.  
 Thus, the precise measurements of the ratio $|V_{ub}/V_{cb}|$ and $\sin 2\beta$ are a crucial test of  our textures.
  
   There is  another  CP violating parameter $J_{CP}$
   in addition to $\sin 2\beta$ and $\delta_{\rm CP}$.
   We also show the predicted $J_{CP}$ versus $|V_{ub}|$ in fig.13,
   where the red dashed lines denote experimental bounds in Eqs. (\ref{data})
  and (\ref{data2}). 
   The some experimental allowed region is excluded in our texture.
   The precise measurements of $|V_{ub}|$ and $J_{CP}$  also provide us a test of our textures.
  
\begin{figure}[h!]
\begin{minipage}[]{0.45\linewidth}
\includegraphics[width=8cm]{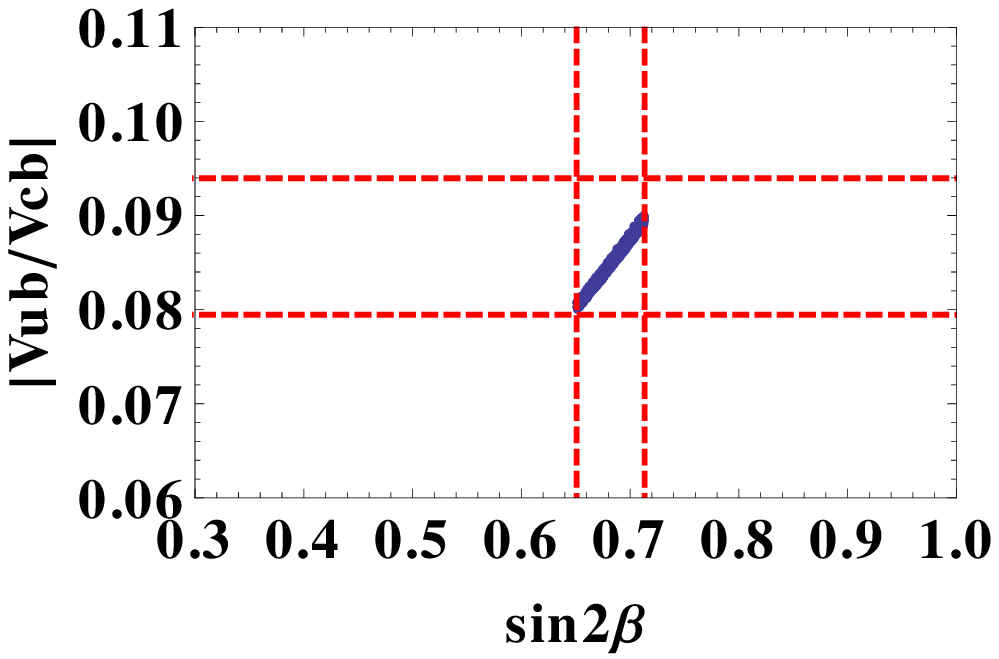}
\caption{The predicted ratio  $|V_{ub}|/|V_{cb}|$ versus $\sin 2\beta$ in $M_{d}^{(1)}$. The red dashed lines denote 
the upper and lower bounds  of the experimental data with $90\%$ C.L.}
\end{minipage}
\hspace{5mm}
\begin{minipage}[]{0.45\linewidth}
\includegraphics[width=7.8cm]{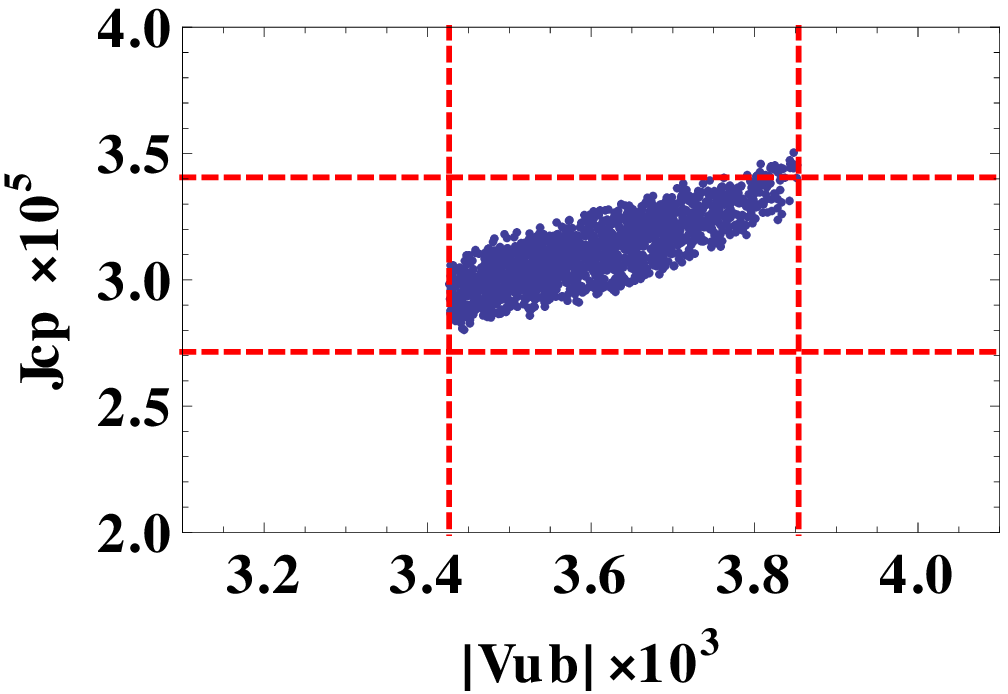}
\caption{The predicted  $J_{CP}$ versus $|V_{ub}|$ in $M_{d}^{(1)}$. The red dashed lines denote 
the upper and lower bounds  of the experimental data with $90\%$ C.L.}
\end{minipage}
\end{figure}

\subsection{Other textures   $M_d^{(k)}(k=2-6, \  11-17)$}

 The predictions of   the mixing angles and the CP violating phase in our thirteen textures
 are classified in the three groups as seen in Table \ref{tab1}.
 The predictions of the first group have been presented for  $M_d^{(1)}$ in the previous subsection.
 Other mass matrices  $M_d^{(k)}(k=2-6, \  11-17)$ can be also studied in the same way.
 We have checked numerically that 
 the Cabibbo angle  is predicted  with a good accuracy by using
the experimental data $\sin 2\beta$, $\theta_{13}$ and $\theta_{23}$ without assuming 
any flavor symmetry.
  We summarize the allowed region of the parameters
 in Table \ref{tab2} and Table \ref{tab3}, where $M_d^{(16)}$ is the exceptional case with
 $c'>d$.
  We have found that only six textures
 are independent   among thirteen textures as shown in Appendix B.
 
\begin{table}[hbtp]
\begin{center}
\begin{tabular}{|c|c|c|c|c|c|c|c|}
\hline 
& $a$ [MeV]& $a'$ [MeV] & $b$ [MeV] &  $c$ [MeV]& $c'$ [GeV]& $d$ [GeV] &$\phi$ [$\circ$]\\
\hline 
$M_{d}^{(1)}$ &$15$-$17.5$&$10$-$15$&$92$-$104$& $78$-$95$&$1.65$-$2.0$ & $2.0$-$2.3$ &$37$-$48$\\
\hline 
$M_{d}^{(2)}$&$15$-$17$ &$2$-$4$ & $94$-$106$&$78$-$95$&$1.65$-$2.0$ &$2.0$-$2.3$ &$40$-$49$\\
\hline 
$M_{d}^{(3)}$&$15$-$17.5$ &$250$-$380$&$92$-$104$&$78$-$95$&$1.65$-$2.0$ &$2.0$-$2.3$ &$37$-$48$\\
\hline 
$M_{d}^{(4)}$&$11$-$14$ &$9$-$17$&$45$-$58$&$115$-$128$&$0.009$-$0.011$ &$2.8$-$2.9$ &$63$-$75$\\
\hline 
$M_{d}^{(5)}$&$11$-$14$ &$2$-$4$&$45$-$58$&$115$-$128$&$0.009$-$0.011$ &$2.8$-$2.9$ &$63$-$75$\\
\hline 
$M_{d}^{(6)}$&$11$-$14$ &$220$-$420$&$45$-$58$&$115$-$128$&$0.009$-$0.011$ &$2.8$-$2.9$ &$63$-$75$\\
\hline 
\end{tabular}
\caption{The allowed regions of parameters for each $M_d^{(k)}(k=1-6)$.}
\label{tab2}
\end{center}
\end{table}
\begin{table}[hbtp]
\begin{center}
\begin{tabular}{|c|c|c|c|c|c|c|c|}
\hline 
& $a$ [MeV]& $a'$ [MeV] & $b$ [MeV] &  $c$ [MeV]& $c'$ [GeV]& $d$ [GeV] &$\phi$ [$\circ$]\\
\hline 
$M_{d}^{(11)}$ &$10$-$12$&$2.5$-$3.5$&$11$-$13$&$125$-$135$&$1.0$-$1.2$ & $2.5$-$2.7$ &$104$-$118$\\
\hline 
$M_{d}^{(12)}$ &$10$-$12$&$11$-$18$&$11$-$13$& $125$-$135$&$1.0$-$1.2$ & $2.5$-$2.7$ &$106$-$120$\\
\hline 
$M_{d}^{(13)}$ &$10$-$12$&$260$-$390$&$11$-$13$& $125$-$135$&$1.0$-$1.2$ & $2.5$-$2.7$ &$104$-$118$\\
\hline 
$M_{d}^{(14)}$&$11$-$14$ &$2$-$4$&$45$-$58$&$115$-$128$&$0.009$-$0.011$&$2.8$-$2.9$ &$63$-$75$\\
\hline 
$M_{d}^{(15)}$&$10$-$12$&$2.5$-$3.5$ &$11$-$13$&$125$-$135$&$1.0$-$1.2$&$2.5$-$2.7$ &$104$-$118$\\
\hline 
$M_{d}^{(16)}$&$2$-$4$&$78$-$95$ &$15$-$17$&$94$-$106$&$2.0$-$2.3$&$1.65$-$2.0$ &$40$-$49$\\
\hline 
$M_{d}^{(17)}$&$15$-$17$ &$2$-$4$&$94$-$106$&$78$-$95$&$1.65$-$2.0$&$2.0$-$2.3$ &$40$-$49$\\
\hline 
\end{tabular}
\caption{The allowed regions of parameters for each $M_d^{(k)}(k=11-17)$.}
\label{tab3}
\end{center}
\end{table}
\begin{figure}[b!]
\begin{minipage}[]{0.45\linewidth}
\includegraphics[width=8cm]{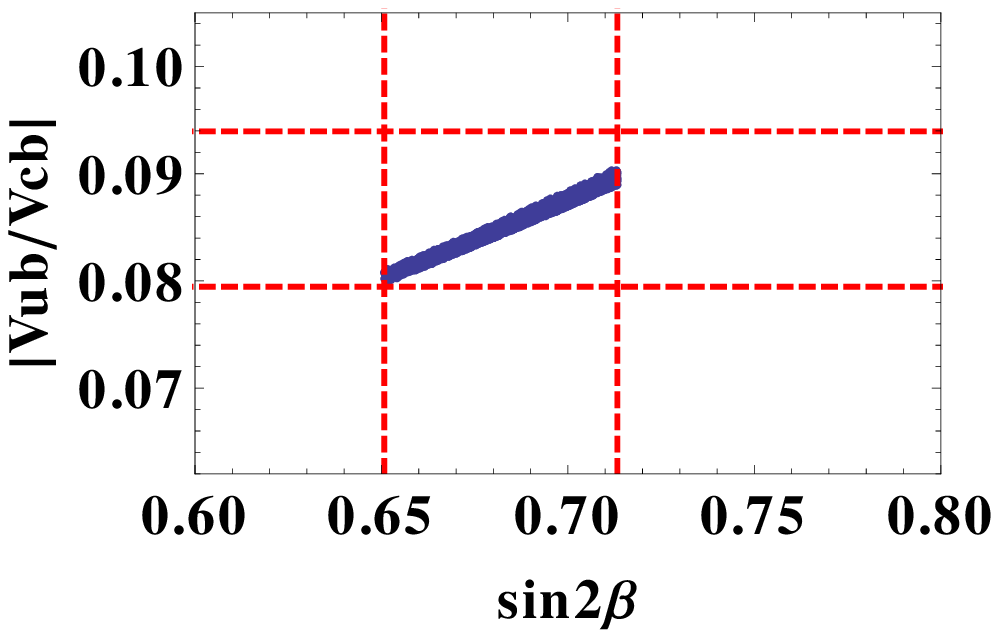}
\caption{The predicted ratio  $|V_{ub}|/|V_{cb}|$ versus $\sin 2\beta$ in $M_{d}^{(5)}$. The red dashed lines denote 
the upper and lower bounds  of the experimental data with $90\%$ C.L. in Eq. (\ref{data}). }
\end{minipage}
\hspace{5mm}
\begin{minipage}[]{0.45\linewidth}
\includegraphics[width=8cm]{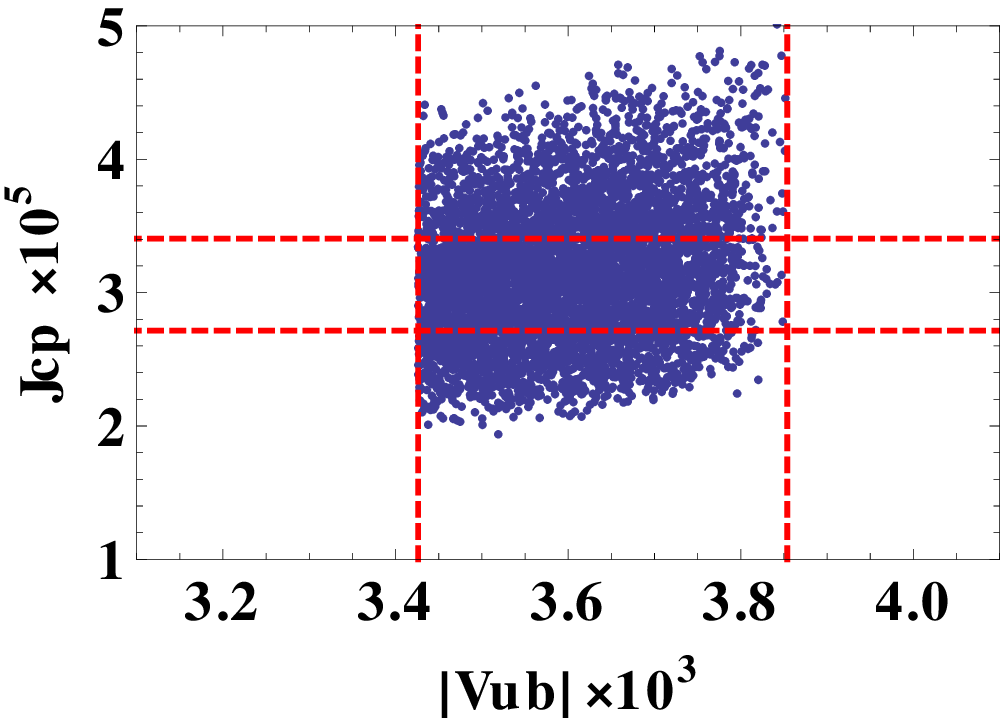}
\caption{The predicted  $J_{CP}$ versus $|V_{ub}|$ in $M_{d}^{(5)}$. The red dashed lines denote 
the upper and lower bounds  of the experimental data with $90\%$ C.L. in Eq. (\ref{data}).}
\end{minipage}
\end{figure}
\begin{figure}[h!]
\begin{minipage}[]{0.45\linewidth}
\includegraphics[width=8cm]{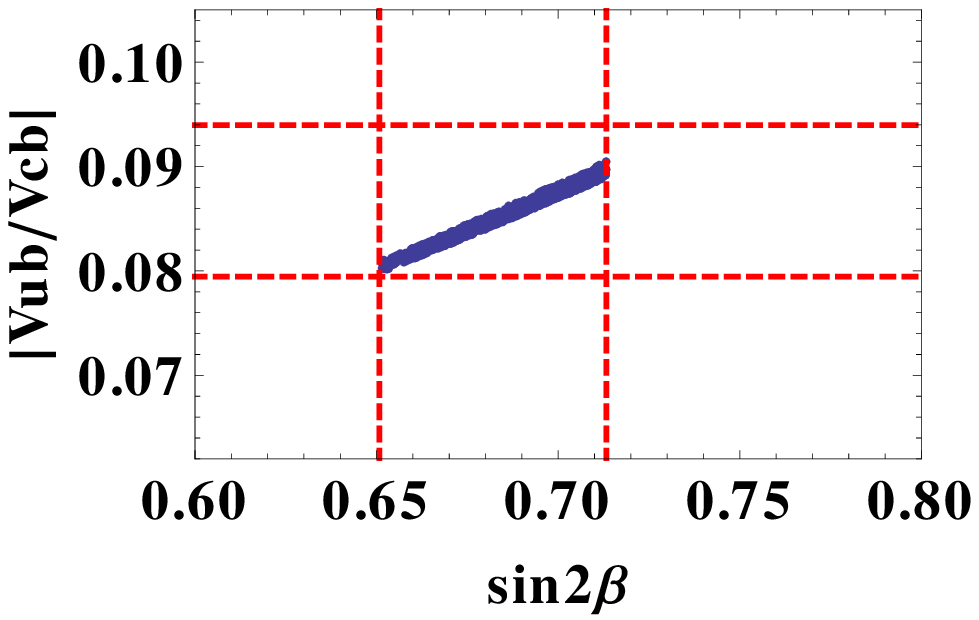}
\caption{The predicted ratio  $|V_{ub}|/|V_{cb}|$ versus $\sin 2\beta$ in $M_{d}^{(11)}$. The red dashed lines denote 
the upper and lower bounds  of the experimental data with $90\%$ C.L. in Eq. (\ref{data}). }
\end{minipage}
\hspace{5mm}
\begin{minipage}[]{0.45\linewidth}
\includegraphics[width=8cm]{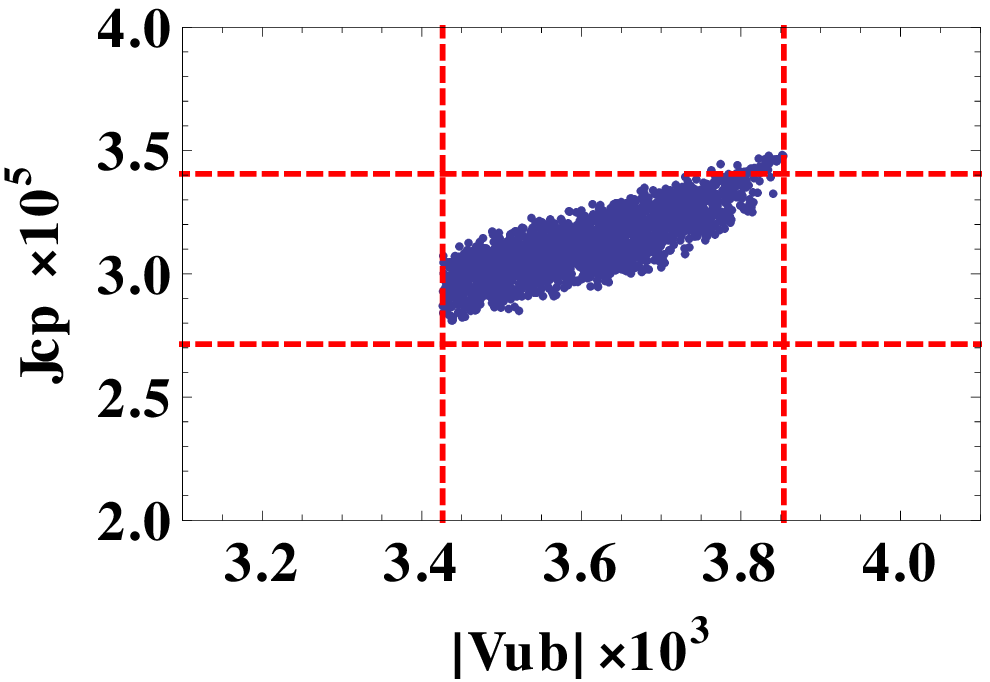}
\caption{The predicted  $J_{CP}$ versus $|V_{ub}|$ in $M_{d}^{(11)}$. The red dashed lines denote 
the upper and lower bounds  of the experimental data with $90\%$ C.L. in Eq. (\ref{data})}
\end{minipage}
\end{figure}
 
 We show here predictions for $M_d^{(5)}$ and $M_d^{(11)}$
 since numerical results are almost the same in the same group in Table 1.
  In fig.14, we show the  predicted ratio $|V_{ub}/V_{cb}|$ versus $\sin 2\beta$ 
 for $M_d^{(5)}$ as a representative of the second  group in Table 1.
 The upper bound of the predicted ratio is also $0.09$, which is the same as the one
   in fig.12 for  $M_d^{(1)}$.
 The prediction is smaller than  the experimental upper bound $0.094$.  
  We  show the  $J_{CP}$ versus $|V_{ub}|$ in fig.15 for $M_d^{(5)}$,
    where the red dashed lines denote experimental bounds in Eqs. (\ref{data})
  and (\ref{data2}). 
  The predicted $J_{CP}$ is  different from the one 
  in fig.13 for  $M_d^{(1)}$.

  In fig.16, we show the predicted ratio $|V_{ub}/V_{cb}|$ versus $\sin 2\beta$ 
 for $M_d^{(11)}$ as a representative of the third  group in Table 1.
  The prediction is  the same as the one
    for  $M_d^{(1)}$ and  $M_d^{(5)}$ as seen in figs.12 and 14.
  We  show the  $J_{CP}$ versus $|V_{ub}|$ for $M_d^{(11)}$  in fig.17.
  The predicted $J_{CP}$ is almost the same as the one in fig.13,
  but different from the one in fig. 15.
   Therefore,  the precise measurements of  $|V_{ub}/V_{cb}|$,
 $J_{CP}$ and $\sin 2\beta$  are important to distinguish the textures.

\section{Discussions and Summary}

In our work,  we consider the minimum number of parameters
of the quark mass matrices  needed  for the  successful CKM mixing angles and the CP violation.
We impose three zeros in the down-quark mass matrix by taking the diagonal  up-quark mass matrix.  The three zeros are maximal zeros  to keep the CP violating phase
 in the quark mass matrix. Then,  there remain six real parameters and one CP violating phase,
which is the minimal number to reproduce the observed data of the down-quark masses and the CKM parameters.
In order to reproduce the bottom-quark mass and the CKM mixing, $V_{us}$ and $V_{cb}$, we take the  $(3,3)$, $(2,3)$,  $(1,2)$ elements of the $3\times 3$ down-quark mass matrix to be non-vanishing.
Therefore, we have $_6 C_3=20$ textures with three zeros. 
 We have found that  the thirteen textures among twenty ones  are viable for the down-quark mass matrix.
 They are classified into three groups in Table 1.
 It is remarked that  
 these textures have freedoms of the unitary transformation
of the  right-handed quarks and some textures are  transformed to other ones.
By using such transformations, we have found that  six  textures are independent among the thirteen textures (see Appendix B).
 
 As a representative of the above six textures, we have discussed the texture $M_{d}^{(1)}$
in details  to see how well the Cabibbo angle is predicted. 
 By imposing the experimental data on $\sin 2\beta$, $\theta_{13}$ and $\theta_{23}$, the Cabibbo angle is predicted to be close to the experimental data.
 We have found that this surprising result remains unchanged in all other viable textures.
 Thus, the  Occam's Razor approach is  very powerful to obtain the successful Cabibbo angle.


After fixing all parameters by using the experimental data of three down-quark masses,  
the three CKM mixing angles and one CP violating phase,
 we have investigated  the correlation between $|V_{ub}/V_{cb}|$ and  $\sin 2\beta$.
 For all textures, the maximal value of the ratio $|V_{ub}/V_{cb}|$ is $0.09$, which is smaller than
 the upper-bound of  the experimental data, $0.094$.
 We have also discussed $J_{CP}$ versus  $|V_{ub}|$.
  The predicted $J_{CP}$ is almost the same among the first and third groups of Table 1, but
   different from the one in the second group.
  The precise data  $|V_{ub}/V_{cb}|$,
 $J_{CP}$ and $\sin 2\beta$  provide us the important test  for our textures.

 Our textures have been analyzed at the electroweak scale in this paper.
The stability of texture zeros of the quark mass matrix has been
examined against the renormalization-group evolution from the GUT scale
to the electroweak scale  by Xing and Zhao \cite{Xing:2015sva}.
They found that   texture zeros of the quark mass matrix are essentially stable against the
evolution.
Thus, we expect that the conclusions derived in this paper do not change much 
even if we consider the textures in 
Eqs.(\ref{downmassmatrixA}) and (\ref{downmassmatrixB}) at the GUT scale.



\vspace{0.5cm}
\noindent
{\bf Acknowledgement}

This work is supported by JSPS Grand-in-Aid for Scientific Research (15K05045;M.T) and
Scientific Research B (No.26287039;TTY).
This work is also supported by World Premier International Research Center Initiative 
(WPI Initiative), MEXT, Japan.

\newpage

\appendix 

\section*{Appendix}


\section{Unfavored down-quark mass matrices}

As discussed in the section 3, the  four textures (seventh-tenth ones) in Eq.(\ref{massmatrixA}) are  excluded by the experimental data. 
The following seventh, eighth and ninth textures are easily shown to be inconsistent with 
the experimental data:
\begin{align} M_d^{(7)}=
\begin{pmatrix}
0 & a & 0 \\
a' & b \ e^{-i\phi}& c \\
c' & 0 & d
\end{pmatrix}  , \quad 
 M_d^{(8)}=
\begin{pmatrix}
a' & a & 0 \\
c' & b \ e^{-i\phi} & c \\
0 & 0 & d
\end{pmatrix} , \quad 
 M_d^{(9)}=
\begin{pmatrix}
0 & a & c' \\
0 & b \ e^{-i\phi} & c \\
0 & a' & d
\end{pmatrix} .
\label{A789}
\end{align}
The both matrices  $M_d^{(7)}  M_d^{(7)\dagger}$ 
and $M_d^{(8)}  M_d^{(8)\dagger}$ give the vanishing $(13), (31)$ elements.
These zeros are consistent  only when $\theta_{13}=0$, $\theta_{23}=90^\circ$
or $\delta_{CP}=0$.
Actually, the Jarlskog invariant $J_{CP}$ of those textures vanishes.
The matrix $M_d^{(9)}$ has one zero eigenvalue, that is, the lightest d-quark is massless.

The last texture in Eq.(\ref{massmatrixA}) is non-trivial one. We parametrize
 the texture as  follows:
\begin{align}
M_d^{(10)}=
\begin{pmatrix}
a' & a & 0 \\
0 & b e^{-i\phi}& c \\
c' & 0 & d
\end{pmatrix}  ,
\end{align}
which gives the CKM elements in the leading order as follows:
\begin{equation}
|V_{us}|\simeq 2\frac{ab}{m_s^2} |\sin\frac{\phi}{2}| \ ,
\quad
|V_{cb}|\simeq \frac{c}{m_b} \ ,
\quad
|V_{ub}|\simeq \frac{a' c'}{m_b^2} \ ,
\quad
\delta_{CP}\simeq \frac{1}{2}(\pi+\phi) \ ,
\end{equation}
where we adopt the approximate relations $a'\sim a$, $c\sim 2b$ and $c'\sim d$.
The $J_{CP}$ is given as
\begin{align}
J_{CP}=\frac{1}{(m_b^2-m_s^2) (m_s^2-m_d^2) (m_b^2-m_d^2) }\ a a' b c c' d \sin\phi   \ .
\end{align}
Therefore, $\sin\phi$ should be positive. Then, $\delta_{CP}$
 is larger than $90^\circ$ or negative, which is completely excluded by the experimental data in Eq.(\ref{data}).
 Actually, we have obtained  $\delta_{CP} \geq 90^\circ$ 
 under the experimental constraint  of $|V_{us}|$, $|V_{cb}|$ and $|V_{ub}|$ numerically.
 
 The  following textures (seventh-tenth ones) in Eq.(\ref{massmatrixB}) are also  excluded by the experimental data:
\begin{align} M_d^{(18)}=
\begin{pmatrix}
a' & a\ e^{i\phi} & 0 \\
0 & 0& c \\
b & c' & d
\end{pmatrix}  , \quad
M_d^{(19)}=
\begin{pmatrix}
0 & a & 0 \\
a' &0& c \\
b &c'\ e^{-i\phi} & d 
\end{pmatrix} ,  \quad
M_d^{(20)}=
\begin{pmatrix}
a' & a & 0 \\
b &0& c\ e^{i\phi} \\
0& c' & d 
\end{pmatrix} .
\end{align}
The both matrices  $M_d^{(18)}  M_d^{(18)\dagger}$ 
and $M_d^{(19)}  M_d^{(19)\dagger}$ give the vanishing $(12), (21)$ elements.
These zeros are consistent  only when  $\theta_{13}=0$, $\theta_{23}=0$ or
$\delta_{CP}=0$.
The Jarlskog invariant $J_{CP}$ of those textures vanishes as well as  $M_d^{(7)}$ 
and $M_d^{(8)}$.
Since   $M_d^{(20)}$ turns to  $M_d^{(10)}$  by the rotation of the right-handed quarks,
 $M_d^{(20)}$ is excluded with  the same reason of $M_d^{(10)}$,   $\delta_{CP}\geq 90^\circ$.
\section{Redundancy of our textures}

Since the CKM matrix is the  flavor mixing among  the left-handed quarks, 
 the  textures in Eqs.(\ref{downmassmatrixA}) and (\ref{downmassmatrixB}) have freedoms of the unitary transformation of the  right-handed quarks.  Under this transformation,  $M_d M_d^\dagger$ is invariant.
We can easily seen that some textures are  transformed to other ones as follows:

\begin{eqnarray}
&&M_d^{(1)} \equiv M_d^{(3)} \ , \qquad M_d^{(2)} \equiv M_d^{(16)} \equiv M_d^{(17)}\ ,  \qquad
 M_d^{(4)} \equiv M_d^{(5)}\equiv M_d^{(14)}  \ , 
\nonumber \\
&& M_d^{(11)} \equiv M_d^{(13)} \equiv M_d^{(15)}  \ , \qquad M_d^{(10)} \equiv M_d^{(20)} \ ,
\end{eqnarray}
where the notation $"\equiv"$ means the equivalent up to   the unitary transformation
of the  right-handed quarks.
In addition,
we can easily see that $M_d^{(7)}$,  $M_d^{(8)}$, $M_d^{(18)}$ and  $M_d^{(19)}$ 
can be  transformed to the textures with four zeros by the right-handed quark rotation.
Therefore, these ones are excluded.
In conclusion, we have  viable six independent textures for the down-quark mass matrix.


\end{document}